\documentclass
[preprint,10pt,nofootinbib,showpacs,nobibnotes,superscriptaddress,balancelastpage,
,titlepage,twocolumn,pre]{revtex4}%
\usepackage{amsfonts}
\usepackage{amsmath}
\usepackage{amssymb}
\usepackage{graphicx}%
\begin{document}
\preprint{ }
\title{Remarks about the thermostatistical description of the HMF model\\Part I: Equilibrium Thermodynamics }
\author{L. Velazquez}
\affiliation{Departamento de F\'{\i}sica, Universidad de Pinar del R\'{\i}o, Mart\'{\i}
270, Esq. 27 de Noviembre, Pinar del R\'{\i}o, Cuba.}
\author{F. Guzman}
\affiliation{Departamento de F\'{\i}sica Nuclear, Instituto Superior de Ciencias y
Tecnolog\'{\i}a Nucleares, Carlos III y Luaces, Plaza, La Habana, Cuba.}
\date{\today}

\begin{abstract}
In this series of papers we\ shall carry out a reconsideration of the
thermodynamical behavior of the called HMF model, a paradigmatic ferromagnetic
toy model exhibiting many features of the real long-range interacting systems.
This first work is devoted to perform the microcanonical description of this
model system: the calculation of microcanonical entropy and some fundamental
thermodynamic observables, the distribution and correlation functions, as well
as the analysis of the thermodynamical stability and the relevant
thermodynamic limit.

\end{abstract}
\pacs{05.20.Gg; 05.20.-y}
\maketitle

\section{Introduction}

The present effort deals with the macroscopic description of a toy model with
a great conceptual significance: the called Hamiltonian Mean Field (HMF) model
\cite{kk,pichon,inaga}. Despite the HMF model is enough amenable for allowing
an accurate numerical and analytical characterization, it exhibits many
features observed in more realistic long-range interacting systems such as:
violent relaxation, persistence of metastable states, slow collisional
relaxation, phase transition, anomalous diffusion, etc
\cite{ant,lat1,lat2,lat3,lat4,zanette,dauxois,yamaguchi,ybd,chava,bouchet1,bouchet2,barre}%
. This reason explains why the present system is considered as a paradigmatic
toy model in the understanding of the thermodynamical and dynamical properties
of the real long-range interacting systems \cite{chava}.

The Hamiltonian of this model system is given by:%

\begin{equation}
H_{N}=\sum_{i=1}^{N}\frac{1}{2I}L_{i}^{2}+\frac{1}{2}g\sum_{i=1}^{N}\sum
_{j=1}^{N}\left[  1-\cos\left(  \theta_{i}-\theta_{j}\right)  \right]  ,
\label{H}%
\end{equation}
where $I$ is the momentum of inertia and $g$ the coupling constant. This is
just an inertial version of the ferromagnetic XY model \cite{stan}, where the
interaction is not restricted to first neighbors but is extended to all
couples of spins. Alternatively, this system could also be interpreted as a
set of particles moving on a circle and interacting via a cosine binary
potential. We shall refer mainly in the present work to the ferromagnetic
interpretation. Besides the total energy $E=H_{N}$, a very important
microscopic observable is the magnetization vector $\mathbf{m=}\left(
\sum_{k=1}^{N}\mathbf{m}_{i}\right)  /N$ where $\mathbf{m}_{i}=\left(
\cos\theta_{i},\sin\theta_{i}\right)  $.

General speaking, the macroscopic characterization of this model system by
using the standard thermostatistical treatment does not significantly differs
from the usual properties of the short-range ferromagnetic models. However,
the understanding of its dynamical features is still an open problem
attracting much attention in the last years
\cite{dauxois,yamaguchi,ybd,chava,bouchet1,bouchet2,barre}.

The present work is just the first part of a series of papers devoted to
perform a re-analysis of the thermodynamical behavior of the HMF model with
the aim to obtain a better understanding of its remarkable dynamical behavior.
As expected, we begin to address the equilibrium thermodynamical properties of
this model system by using the microcanonical description. As elsewhere
discussed, the microcanonical ensemble is just a \textit{dynamical ensemble}
associated to the ergodic character of the many-body nonlinear Hamiltonian
dynamics. Thus, the using of this statistical ensemble allows to keep a close
relationship with the dynamical behavior of the HMF model studied by means of
numerical microcanonical simulations. Besides the entropy and some relevant
thermodynamical parameters, the magnetic susceptibility, the particles
distribution and the two-body correlation function will be also obtained. Two
novel aspects accounted for at the end of the present discussion are the
analysis of thermodynamical stability and the relevance of the thermodynamic limit.

\section{\label{thermo}Equilibrium Thermodynamics}

The thermodynamical description of the HMF model by using the mean field
approximation was\ considered in refs.\cite{chava,barre}. We shall perform in
this section the microcanonical description directly working on the
\textit{N}-body phase space. As usual, the microcanonical expectation value of
any microscopic observable $A\left(  X\right)  $ is obtained from the expression:%

\begin{equation}
A_{m}=\frac{1}{\Omega}Sp\left[  A\left(  X\right)  \delta\left(
E-H_{N}\right)  \right]  ,
\end{equation}
where $\Omega$ is microcanonical states density, $\Omega=Sp\left[
\delta\left(  E-H_{N}\right)  \right]  $, while $Sp\left[  Q\right]  $
represents the N-body phase space integration:%

\begin{equation}
Sp\left[  Q\right]  \equiv\int dXQ\left(  X\right)  =\frac{1}{N!}\int
\frac{d^{N}\theta d^{N}L}{\left(  2\pi\hslash\right)  ^{N}}Q\left(
\theta,L\right)  .
\end{equation}
where the factorial term considers the particles identity.

\subsection{Thermodynamical parameters\label{bolt}}

The microcanonical states density $\Omega$ is calculated as follows:%

\begin{equation}
\Omega\left(  E,N;I,g\right)  =\frac{1}{N!}\int\frac{d^{N}\theta d^{N}%
L}{\left(  2\pi\hslash\right)  ^{N}}\delta\left[  E-H_{N}\left(
\theta,L;I,g\right)  \right]  .
\end{equation}
The integration by $d^{N}L$ yields:%

\begin{equation}
\frac{1}{N!\Gamma\left(  \frac{N}{2}\right)  }\left(  \frac{2\pi I}%
{\hslash^{2}}\right)  ^{\frac{1}{2}N}\int\frac{d^{N}\theta}{\left(
2\pi\right)  ^{N}}\left[  E-V_{N}\left(  \theta;g\right)  \right]  ^{\frac
{1}{2}N-1}\text{.}%
\end{equation}
The accessible volume $W$ is expressed in a dimensionless form as $W=\Omega
g/2$. This last equation is rewritten when $N$ tends to infinity as follows:%

\begin{equation}
W\simeq\left(  \frac{2\pi e^{3}Ig}{N\hslash^{2}}\right)  ^{\frac{1}{2}N}\int
d^{2}\mathbf{m}\left[  \mathbf{m}^{2}+2u-1\right]  ^{\frac{1}{2}N-1}f\left(
\mathbf{m};N\right)  , \label{WW}%
\end{equation}
where the dimensionless energy $u=E/gN^{2}$ and the magnetization distribution
function $f\left(  \mathbf{m};N\right)  $:%

\begin{equation}
f\left(  \mathbf{m};N\right)  =N^{2}\int\frac{d^{N}\theta}{\left(
2\pi\right)  ^{N}}\delta\left[  N\mathbf{m}-\overset{N}{\underset{k=1}{\sum}%
}\mathbf{m}_{k}\right]  , \label{dist}%
\end{equation}
were introduced. The distribution function (\ref{dist}) can be rephrased by
using the Fourier representation of the Dirac delta function as follows:%

\begin{equation}
N^{2}\int\frac{d^{2}\mathbf{k}}{\left(  2\pi\right)  ^{2}}\exp\left(
iN\mathbf{K}\cdot\mathbf{m}\right)  \left[  I_{0}\left(  -i\sqrt
{\mathbf{K}^{2}}\right)  \right]  ^{N}, \label{compleja}%
\end{equation}
where $\mathbf{K}=\mathbf{k}+i\mathbf{x}$, being $\mathbf{x}$ a real
bidimensional vector and $I_{n}\left(  z\right)  $ the modified Bessel
function of \textit{n}-th order:%

\begin{equation}
I_{n}\left(  z\right)  =\frac{1}{2\pi}\int_{-\pi}^{+\pi}\exp\left(
z\cos\theta\right)  \cos\left(  n\theta\right)  d\theta, \label{bessel}%
\end{equation}
which satisfies the recurrence relations:%

\begin{align}
I_{n+1}\left(  z\right)  -I_{n-1}\left(  z\right)   &  =-\frac{2n}{z}%
I_{n}\left(  z\right)  ,\label{recu1}\\
I_{n-1}\left(  z\right)  +I_{n+1}\left(  z\right)   &  =2\frac{dI_{n}\left(
z\right)  }{dz}. \label{recu2}%
\end{align}

The main contribution of the integral when $N$ tends to infinity is obtained
by using the steepest descend method:%

\begin{align}
f\left(  \mathbf{m};N\right)   &  \simeq\exp\left\{  -N\left[  xm-\ln
I_{0}\left(  x\right)  \right]  -\right. \nonumber\\
&  \left.  -\frac{1}{2}\ln\left[  \left(  \frac{2\pi}{N}\right)  ^{2}%
\kappa_{1}\left(  x\right)  \kappa_{2}\left(  x\right)  \right]  \right\}  ,
\label{func}%
\end{align}
where the functions $\kappa_{1}\left(  x\right)  $ and $\kappa_{2}\left(
x\right)  $ are given by:%
\begin{equation}
\kappa_{1}\left(  x\right)  =\frac{2m\left(  x\right)  }{x}+\frac{d}%
{dx}m\left(  x\right)  >0,~\kappa_{2}=\frac{m\left(  x\right)  }{x}>0.
\end{equation}
being $m=\left\vert \mathbf{m}\right\vert $ the magnetization modulus related
to $x=\left\vert \mathbf{x}\right\vert $ by:%

\begin{equation}
m=m\left(  x\right)  =\frac{I_{1}\left(  x\right)  }{I_{0}\left(  x\right)  }.
\label{mx}%
\end{equation}

The validity of the above approximation is ensured by the \textit{positivity}
of the argument of the logarithmic term in Eq.(\ref{func}). The derivation of
this expression appears in the appendix \ref{expan}. Notice that the
integration function of Eq.(\ref{WW}) only depends on the modulus of
$\mathbf{m}$, and therefore, the microcanonical expectation value of
$\mathbf{m}$ identically vanishes as a consequence of the nonexistence of a
preferential direction of this vector. Nevertheless, the expectation value of
$\mathbf{m}$ can differ from zero as a consequence of the occurrence of a
spontaneous symmetry breaking (see in the next subsection).

\begin{figure}
[t]
\begin{center}
\includegraphics[
height=2.8106in,
width=3.5129in
]%
{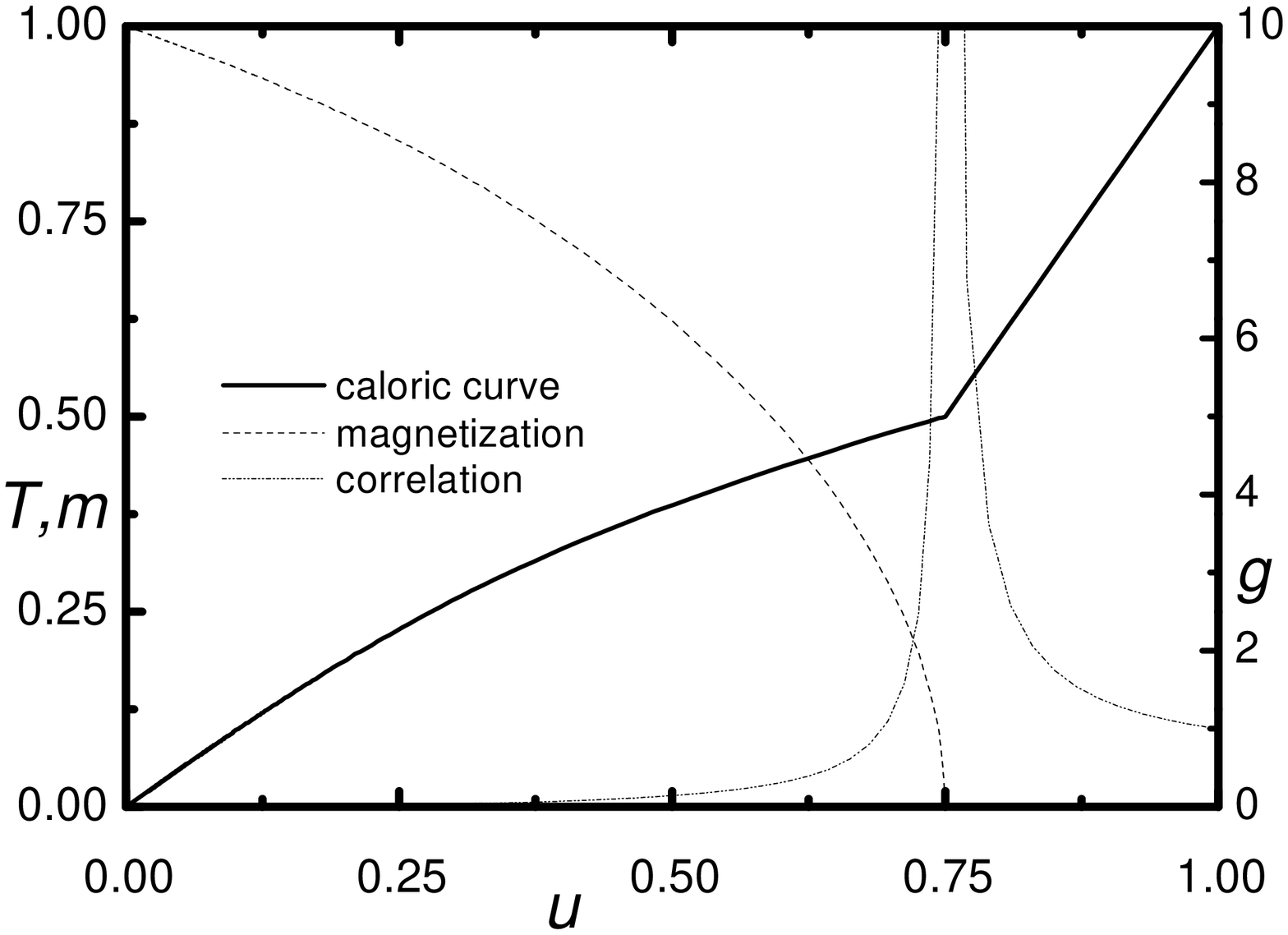}%
\caption{Microcanonical description of the HMF model in the thermodynamic
limit: the caloric $T\left(  u\right)  $ (solid line) and magnetization
$m\left(  u\right)  $ (dash line) curves clearly reveal the existence of a
second-order phase transition at $u_{c}=0.75$ with $T_{c}=0.5$. The divergence
of the correlation function $g\left(  u\right)  $ (dash-dot-dot line)
manifests the existence of a long-range order at the critical point $u_{c}$.}%
\label{micro.eps}%
\end{center}
\end{figure}

The main contribution to the entropy per particle $s\left(  u,N;I,g\right)
=\ln W/N$ when $N$ tends to infinity can be obtained by using again the
steepest descend method as follows:%

\begin{align}
s\left(  u,N;I,g\right)   &  =s_{0}+\max_{x}\left\{  \frac{1}{2}\ln\left[
m^{2}\left(  x\right)  +\kappa\right]  \right. \nonumber\\
&  \left.  -xm\left(  x\right)  +\ln I_{0}\left(  x\right)  \right\}
+O(\frac{1}{N}), \label{s per part}%
\end{align}
where $\kappa=2u-1$ and the additive constant $s_{0}=%
\frac12
\ln\left(  2\pi e^{3}Ig/N\right)  $. The stationary condition is given by:%

\begin{equation}
\left[  \frac{m\left(  x\right)  }{x}-m^{2}\left(  x\right)  -\kappa\right]
\frac{x}{m^{2}\left(  x\right)  +\kappa}\frac{d}{dx}m\left(  x\right)
=0\text{,} \label{con}%
\end{equation}
arriving in this way to the magnetization modulus \textit{versus} energy dependence:%

\begin{equation}%
\begin{array}
[c]{c}%
\left.
\begin{array}
[c]{c}%
m=m(x)\\
u=\frac{1}{2}m\left(  x\right)  /x+\frac{1}{2}\left[  1-m^{2}\left(  x\right)
\right]
\end{array}
\right\}  \text{with }x\in\left[  0,\infty\right)  ,\\
\text{and }m=0\text{ \ if \ }u\geq u_{c}=0.75.
\end{array}
\label{umx}%
\end{equation}
The caloric curve is obtained from the canonical parameter $\eta
=gN\beta=\partial s\left(  u,N;I,g\right)  /\partial u$:%

\begin{equation}
\eta=\frac{1}{T}=\left\{
\begin{tabular}
[c]{ll}%
$x/m\left(  x\right)  $ & with $0\leq u<u_{c},$\\
$1/\left(  2u-1\right)  $ & otherwise,
\end{tabular}
\ \ \ \ \ \ \ \ \ \right.  \label{bmx}%
\end{equation}
where $T$ is the dimensionless temperature. The dependences shown in
FIG.\ref{micro.eps} confirm the existence of a continuous (second-order) phase
transition at the critical energy $u_{c}=0.75$ with critical temperature
$T_{c}=0.5$: from a ferromagnetic phase when $u<u_{c}$, towards a paramagnetic
one with $u>u_{c}$.%

Since $m\left(  x\right)  $ drops to zero when $u\rightarrow u_{c}^{-}$, the
use of the power expansion:%

\begin{equation}
m\left(  x\right)  =\frac{1}{2}x-\frac{1}{16}x^{3}+\frac{1}{96}x^{5}%
+..,\label{mxexp}%
\end{equation}
allows to express $x$ in terms of $u$ as follows:%

\begin{equation}
x^{2}=\frac{32}{5}\left(  u_{c}-u\right)  +\allowbreak\frac{7}{30}\left(
\frac{32}{5}\right)  ^{2}\left(  u_{c}-u\right)  ^{2}+..,
\end{equation}
and therefore, the temperature and magnetization dependences in the
neighborhood of the critical point with $u<u_{c}$ are given by:%

\begin{align}
T\left(  u\right)   &  =\frac{1}{2}-\frac{2}{5}\left(  u_{c}-u\right)
-\frac{64}{375}\left(  u_{c}-u\right)  ^{2}+..,\\
m\left(  u\right)   &  =2\sqrt{\frac{2}{5}\left(  u_{c}-u\right)  }\left(
1-\frac{4}{75}\left(  u_{c}-u\right)  +..\right)  .
\end{align}
Thus, the heat capacity $C\left(  u\right)  =\left(  dT/du\right)  ^{-1}$
undergoes a discontinuity at $u_{c}$:%

\begin{equation}
C\left(  u_{c}^{+}\right)  -C\left(  u_{c}^{-}\right)  =2,
\end{equation}
being $C\left(  u\right)  =1/2$ when $u>u_{c}$.

The results obtained so far are in a total agreement with the one derived from
the mean field approximation carried out in ref.\cite{chava}. As already
evidenced, the thermodynamical features of the HMF model does not essentially
differ from the ones exhibited by other ferromagnetic systems in the mean
field approximation \cite{stan}. Therefore, the solution above provided should
lost its validity in the neighborhood of the critical point $u_{c}$
\cite{chava,chava2,chava3}. In order to show this fact, let us consider the
second derivative of the maximization problem (\ref{s per part}) which allows
to check the stability of the stationary solution (\ref{umx}):%
\begin{equation}
h\left(  x;u\right)  =\left(  m^{\prime}\left(  x\right)  \frac{\kappa
-m^{2}\left(  x\right)  }{\left(  m^{2}\left(  x\right)  +\kappa\right)  ^{2}%
}-1\right)  m^{\prime}\left(  x\right)  .\label{hh}%
\end{equation}
We have used the recurrence relations (\ref{recu1}) and (\ref{recu2}) in order
to express the first derivative of the magnetization as $m^{\prime}\left(
x\right)  =1-m\left(  x\right)  /x-m^{2}\left(  x\right)  $. The stationary
condition (\ref{con}) allows to substitute the relation $\kappa=\kappa\left(
x\right)  \equiv m\left(  x\right)  /x-m^{2}\left(  x\right)  $ into
(\ref{hh}). Thus, the second derivative (\ref{hh}) is negative everywhere with
the exception of the stationary solution $x=0$ corresponding to the
paramagnetic phase:
\begin{equation}
h\left(  x=0;u\right)  \simeq-\frac{u-u_{c}}{2u-1},
\end{equation}
which turns unstable when $u<u_{c}$. The function $h\left(  x;u\right)  $ is
related to the\ average square dispersion of the magnetization throughout the
expression:
\begin{equation}
\left\langle \delta\mathbf{m}^{2}\right\rangle \simeq\left\langle m^{\prime
}\left(  x\right)  \delta x\right\rangle ^{2}=-\frac{\left(  m^{\prime}\left(
x\right)  \right)  ^{2}}{Nh\left(  x;u\right)  },
\end{equation}
which diverges at the critical point of the second-order phase transition
$u_{c}$, yielding:
\begin{equation}
\left\langle \delta\mathbf{m}^{2}\right\rangle =\frac{1}{4N}\frac
{2u-1}{u-u_{c}},
\end{equation}
when $u>u_{c}$. Since the average magnetization square dispersion is directly
related to the correlation functions $g_{ij}=\left\langle \mathbf{m}_{i}%
\cdot\mathbf{m}_{j}\right\rangle -\left\langle \mathbf{m}_{i}\right\rangle
\cdot\left\langle \mathbf{m}_{i}\right\rangle $ via the formula:%
\begin{equation}
N\left\langle \delta\mathbf{m}^{2}\right\rangle \equiv\frac{1}{N}\sum
_{ij}g_{ij}=g\left(  u\right)  ,
\end{equation}
the divergence of $\left\langle \delta\mathbf{m}^{2}\right\rangle $ implies
the existence of a \textit{long-range order} at $u_{c}$ \cite{gallavotti} (see
also in FIG.\ref{micro.eps}). It is well-known that the existence of such a
phenomenon significantly modifies the behavior of the thermodynamical
quantities close to the critical point $u_{c}$.

\subsection{\label{magnet}Magnetic susceptibility}

\begin{figure}
[t]
\begin{center}
\includegraphics[
height=2.8106in,
width=3.5129in
]%
{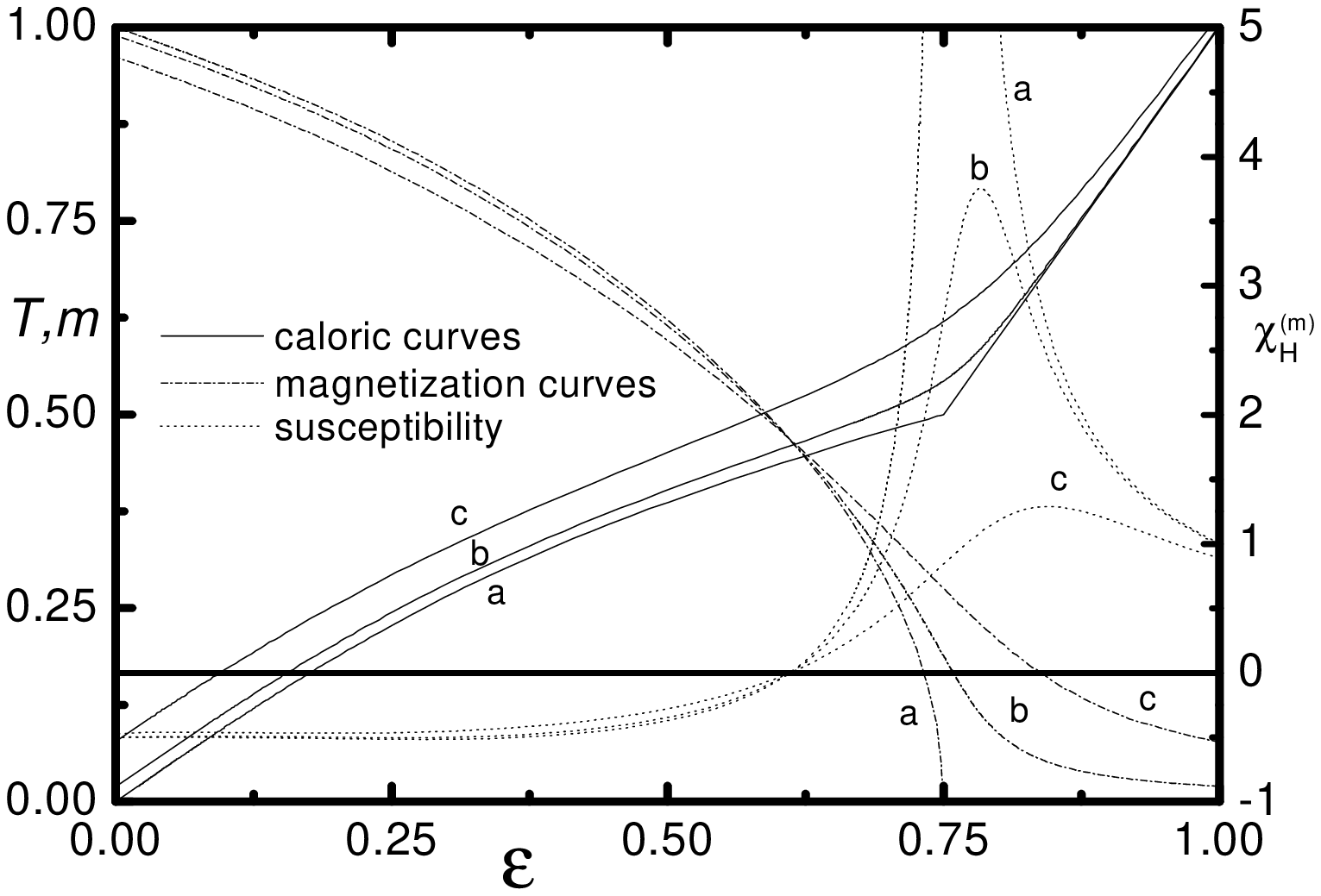}%
\caption{Microcanonical description of the HMF model under the incidence of a
magnetic field: caloric (solid lines) and magnetization curves (dash-dot), as
well as the microcanonical susceptibility (dot lines) for (a) with $H=0$, (b)
with $H=0.02$, and (c) with $H=0.08$. }%
\label{microh.eps}%
\end{center}
\end{figure}

The Hamiltonian of the HMF model is invariant under the translation operation
$\theta_{k}\rightarrow\theta_{k}+\psi$, which is equivalent to the $U\left(
1\right)  $ rotational symmetry acting on the rotator directions, whose
existence implies the vanishing of the expectation value of\ the magnetization
$\mathbf{m}$. The $U\left(  1\right)  $ symmetry is broken by modifying the
Hamiltonian (\ref{H}) with the incidence of an external magnetic field
$\mathbf{H}$ as follows:%

\begin{equation}
H_{N}\rightarrow H_{N}^{\ast}=H_{N}-gN\sum_{k}\mathbf{H}\cdot\mathbf{m}%
_{k}\mathbf{.}\label{HB}%
\end{equation}
\ The vector $\mathbf{H}$ introduces now a preferential direction for the
average magnetization, which leads to a modification of Eq.(\ref{WW}) as follows:%

\begin{equation}
W\propto\int d^{2}\mathbf{m}~f\left(  m;N\right)  \left(  m^{2}+\kappa
+2\mathbf{H}\cdot\mathbf{m}\right)  ^{\frac{1}{2}N-1},
\end{equation}
as well as the following dependences of the caloric and magnetization curves:%

\begin{align}
\varepsilon\left(  x;H\right)   &  =\frac{m\left(  x\right)  +H}{2x}+\frac
{1}{2}\left[  1-m^{2}\left(  x\right)  \right]  -hm\left(  x\right)
,\label{ubx}\\
\eta &  =\frac{1}{T}=\frac{x}{m\left(  x\right)  +H},~m\left(  x\right)
=\frac{I_{1}\left(  x\right)  }{I_{0}\left(  x\right)  }.\label{beta2}%
\end{align}
where $-H\leq\varepsilon\leq+\infty$, $0\leq T\leq+\infty$, $0\leq m\leq1$
when $x$ goes from the infinity to zero, being $H=\left\vert \mathbf{H}%
\right\vert $. Notice that we have distinguished between the dimensionless
total energy $\varepsilon=u-Hm$ associated to the modified Hamiltonian
(\ref{HB}) and the dimensionless internal energy $u$ of the Hamiltonian
(\ref{H}). These curves are represented in FIG.\ref{microh.eps} by considering
different values of the external field $H$. Thus, the system exhibits a
nonvanishing magnetization $\mathbf{m}$ for every finite energy $u$, as well
as there is now a smooth dependence between $T$ and $\varepsilon$.%

The caloric and magnetization curves (\ref{umx}) and (\ref{bmx}) are obtained
from (\ref{ubx}) and (\ref{beta2}) when $\mathbf{H}\rightarrow0$. The
expectation value of the magnetization drops to zero with the vanishing of $H$
when $u>u_{c}$, and therefore, the system recovers the original $U\left(
1\right)  $ symmetry. However, a nonzero magnetization along the direction of
the external field survives\ when the vanishing of $\mathbf{H}$ is carried out
by keeping fixed its orientation. It means that the initial symmetry $U\left(
1\right)  $ is spontaneously broken when $u<u_{c}$.

Let us now obtain the magnetization dependence for low values of $H$ when
$\varepsilon>u_{c}$. Since $m\left(  x\right)  $ and $x$ simultaneously vanish
when $H\rightarrow0$, we are able to use the power expansion (\ref{mxexp}) in
order to obtain the power expansion of the $H$ in terms of $x$ starting from
Eq.(\ref{ubx}):
\begin{equation}
H=2x\left(  \varepsilon-u_{c}\right)  +x^{3}\left(  2\varepsilon-\frac{19}%
{16}\right)  +O\left(  x^{4}\right)  \allowbreak,
\end{equation}
whose inversion leads to the\ expression:%

\begin{equation}
x=\frac{1}{2\left(  \varepsilon-u_{c}\right)  }H-\frac{1}{4^{4}}%
\frac{32\varepsilon-19}{\left(  \varepsilon-u_{c}\right)  ^{4}}H^{3}+O\left(
H^{4}\right)  .
\end{equation}
We finally obtain the $m$ \textit{versus} $H$ dependence by substituting this
latter expansion into Eq.(\ref{mxexp}), yielding:%

\begin{equation}
m\left(  \varepsilon;H\right)  =\frac{1}{4\left(  \varepsilon-\varepsilon
_{c}\right)  }H-\frac{36\varepsilon-22}{4^{4}\left(  \varepsilon-u_{c}\right)
^{4}}H^{3}+O\left(  H^{4}\right)  .
\end{equation}
$\allowbreak$Thus, the microcanonical susceptibility $\chi_{H}^{\left(
m\right)  }$ when $u>u_{c}$ is given by:%
\begin{equation}
\chi_{H}^{\left(  m\right)  }=\left(  \frac{\partial m}{\partial H}\right)
_{\varepsilon}=\frac{1}{4\left(  \varepsilon-u_{c}\right)  }-\frac{3}{4^{4}%
}\frac{36\varepsilon-22}{\left(  \varepsilon-u_{c}\right)  ^{4}}H^{2}+O\left(
H^{3}\right)  , \label{qumuc}%
\end{equation}
which diverges at the critical energy $u_{c}$.

The dependence of the magnetization\ $m$ on the external field $H$ when
$\varepsilon<u_{c}$ can be represented as $m=m\left(  x_{H}\right)  $, where
$x_{H}=x\left(  \varepsilon;H\right)  $ is the solution of the problem:%

\begin{equation}
\varepsilon\left[  x_{H};H\right]  =\varepsilon<u_{c}\text{.}\label{ub}%
\end{equation}
being $\varepsilon\left(  x,H\right)  $ the function defined by Eq.(\ref{ubx}%
). The solution for small values of $H$ can be obtained by using the power
expansions:%
\begin{align}
x_{H} &  =x+c_{1}\left(  x\right)  H+c_{2}\left(  x\right)  H^{2}%
+..,\label{xb}\\
m_{H} &  =m\left(  x_{H}\right)  =m\left(  x\right)  +\chi_{1}\left(
x\right)  H+\chi_{2}\left(  x\right)  H^{2}+..,\label{mb}%
\end{align}
being $\varepsilon\left(  x,0\right)  =\varepsilon$ and $\chi_{1}\left(
x\right)  =m^{\prime}\left(  x\right)  c_{1}\left(  x\right)  $, $\chi
_{2}\left(  x\right)  =\frac{1}{2}m^{\prime\prime}\left(  x\right)  c_{1}%
^{2}\left(  x\right)  +m^{\prime}\left(  x\right)  c_{2}\left(  x\right)  $,
and so on. The substitution of Eqs.(\ref{xb}) and (\ref{mb}) into (\ref{ub})
yields the following results for the function $\chi_{1}$:%

\begin{equation}
\chi_{1}\left(  x\right)  =\frac{m^{\prime}\left(  x\right)  \left(  m\left(
x\right)  -\frac{1}{2x}\right)  }{\left(  -m\left(  x\right)  m^{\prime
}\left(  x\right)  -\frac{1}{2}\frac{m\left(  x\right)  }{x^{2}}+\frac{1}%
{2x}m^{\prime}\left(  x\right)  \right)  },
\end{equation}
and $\chi_{2}=p_{1}\left(  x\right)  /p_{2}\left(  x\right)  $, being:%

\begin{align}
p_{1}\left(  x\right)   &  =\chi_{1}\left(  x\right)  +\frac{1}{2}\chi_{1}%
^{2}\left(  x\right)  -\frac{m\left(  x\right)  }{x^{3}}\left(  \frac{\chi
_{1}\left(  x\right)  }{m^{\prime}\left(  x\right)  }\right)  ^{2}+\nonumber\\
&  +\frac{1}{2x^{2}}\frac{\chi_{1}\left(  x\right)  }{m^{\prime}\left(
x\right)  }\left(  \chi_{1}\left(  x\right)  +1\right)  ,
\end{align}

\begin{equation}
p_{2}\left(  x\right)  =-m\left(  x\right)  +\frac{1}{2x}-\frac{1}{2x^{2}%
}\frac{m\left(  x\right)  }{m^{\prime}\left(  x\right)  }.
\end{equation}

The function $\chi_{1}\left(  x\right)  $ represents the zero-order
approximation of the power expansion for the microcanonical susceptibility
$\chi_{H}^{\left(  m\right)  }\left(  x,H\right)  $ in terms of the magnetic
field $H$. It is possible to show that $\chi_{H}^{\left(  m\right)  }%
\simeq\chi_{1}$ also diverges at $u_{c}$ as follows:%

\begin{equation}
\chi_{H}^{\left(  m\right)  }\simeq\frac{1}{8}\allowbreak\frac{1}{\left(
u_{c}-\varepsilon\right)  }, \label{qumeuc}%
\end{equation}
when $\varepsilon<u_{c}$.

The above estimations of the microcanonical susceptibility $\chi_{H}^{\left(
m\right)  }$ are only applicable when $\left\vert \chi_{1}H\right\vert <<1$.
In general, the magnetic susceptibility can be obtained from the formula:%

\begin{equation}
\chi_{H}^{\left(  m\right)  }=m^{\prime}\left(  x\right)  \left(
\frac{\partial x}{\partial H}\right)  _{\varepsilon}.
\end{equation}
which yields:%

\begin{equation}
\chi_{H}^{\left(  m\right)  }=m^{\prime}\left(  m-\frac{1}{2x}\right)  \left(
\frac{m^{\prime}}{2x}-\frac{m+H}{2x^{2}}-\left(  m+H\right)  m^{\prime
}\right)  ^{-1}.
\end{equation}

The microcanonical susceptibility $\chi_{H}^{\left(  m\right)  }$\ was also
represented in FIG.\ref{microh.eps} for different values of the external field
$H$. Notice that $\chi_{H}^{\left(  m\right)  }$ remains almost constant at
negative value $\chi_{H}^{\left(  m\right)  }\simeq-0.46$ when $\varepsilon
<u_{1}=0.5$. The susceptibility begins to grow beyond the point $u_{1}$ and
vanishes at $u_{2}\simeq0.61$ independently of the value of the external field
$H$, a behavior provoked by the presence of the factor $m\left(  x\right)
-0.5/x$. The system exhibits a large susceptibility under the influence of a
very small external magnetic field in the energetic range $\left(  u_{2}%
,u_{c}\right)  $, which is related to the existence of the second-order phase
transition at $u_{c}$.

It worth to remark that the microcanonical susceptibility $\chi_{H}^{\left(
m\right)  }$ also admits \textit{negative values} when $\varepsilon<u_{2}$.
According to the well-known theorem derived from the canonical description:%
\begin{equation}
\chi_{H}^{\left(  c\right)  }=\frac{\partial\left\langle M\right\rangle _{c}%
}{\partial H}=\beta G_{c},\label{sucep.c}%
\end{equation}
being $G_{c}=\left\langle \Delta\mathbf{M}^{2}\right\rangle _{c}$ the average
square dispersion of the magnetization within this ensemble, the magnetic
susceptibility should be nonnegative. Actually, the microcanonical
susceptibility $\chi_{H}^{\left(  m\right)  }$ obtained in this subsection
provides a measure of the magnetic sensibility of the system at constant
energy instead of at constant temperature. It can be shown that the
microcanonical counterpart of the identity (\ref{sucep.c}) in the
thermodynamic limit \cite{vel.geo} is given by:
\begin{equation}
\chi_{H}^{\left(  m\right)  }=\frac{\partial\left\langle M\right\rangle _{m}%
}{\partial H}=\beta G_{m}+\left\langle M\right\rangle _{m}\frac{\partial
\left\langle M\right\rangle _{m}}{\partial E},\label{sucep.m}%
\end{equation}
which clarifies that the negative values of $\chi_{H}^{\left(  m\right)  }$
come from the term $M\partial M/\partial E$ since usually $\partial M/\partial
E<0$ in the ferromagnetic phase.

\subsection{Distribution functions\label{disfunc}}

Let us now obtain the microcanonical n-body distribution functions of this
model: $F_{m}^{\left(  n\right)  }\left[  y\right]  =F_{m}^{\left(  n\right)
}\left(  y_{1},\ldots y_{n}\right)  $:%

\begin{equation}
F_{m}^{\left(  n\right)  }\left[  y\right]  =\frac{1}{\Omega}Sp\left[
\delta\left(  y_{1}-x_{1}\right)  \ldots\delta\left(  y_{n}-x_{n}\right)
\delta\left(  E-H_{N}\right)  \right]  , \label{f1}%
\end{equation}
where $x_{k}=\left(  \theta_{k},L_{k}\right)  $. We shall assume that in the
ferromagnetic phase there is a nonvanishing magnetization $\mathbf{m=}\left(
m,0\right)  $, which could be obtained by considering the Hamiltonian
(\ref{HB}) with \textbf{$H$}$=\left(  H,0\right)  $ and sending $H$ to zero.

The state density $\Omega$ can be rephrased in a functional form $\mathcal{F}$
as follows:%

\begin{equation}
\Omega\propto\mathcal{F}\left[  \psi,N\right]  =\int d^{2}\mathbf{m~}f\left(
\mathbf{m},N\right)  \left[  \psi\right]  ^{\frac{1}{2}N-1},
\end{equation}
where\ $\psi=\psi\left(  \mathbf{m};N\right)  =2u-1+\mathbf{m}^{2}$ (see in
Eq.(\ref{WW})). It is easy to see that the n-body distribution functions can
be also rephrased by using this same functional form as follows:%

\begin{equation}
F_{m}^{\left(  n\right)  }\propto\frac{\mathcal{F}\left[  \psi+\delta\psi
_{n};N-n\right]  }{\mathcal{F}\left[  \psi,N\right]  }, \label{f1w}%
\end{equation}
where $\delta\psi_{n}$ is given by:%

\begin{equation}
\delta\psi_{n}=-\frac{2}{N}\left(  \sum_{k=1}^{n}\tilde{\varepsilon}%
_{k}\right)  +\frac{1}{N^{2}}\left(  \sum_{k=1}^{n}\delta\mathbf{m}%
_{k}\right)  ^{2},
\end{equation}
being $\mathbf{m}_{k}=\mathbf{m}\left(  \theta_{k}\right)  =\left(  \cos
\theta_{k},\sin\theta_{k}\right)  $ the magnetization vector, $\delta
\mathbf{m}_{k}=\mathbf{m}_{k}-\mathbf{m}$, the corresponding dispersion, and
$\tilde{\varepsilon}_{k}-\mathbf{m}^{2}=\frac{1}{2}p_{k}^{2}-\mathbf{m}%
\cdot\mathbf{m}_{k}\equiv\varepsilon_{k}$ the energy of the \textit{k-th}
rotator. We also consider for convenience the dimensionless momentum
$p_{k}=L_{k}/L_{0}$ by introducing the characteristic unit $L_{0}=\sqrt{IgN}$.

The expression (\ref{f1w}) suggests to perform a perturbative expansion in
power series of $1/N$ of the functional $\ln\mathcal{F}\left(  \psi+\delta
\psi_{n};N-n\right)  $ by taking into account the Gaussian localization of
these integrals when $N$ tends to infinity. We obtain after some algebra the
following result:
\begin{equation}
F_{m}^{\left(  n\right)  }\left[  y\right]  \propto\int d^{2}\mathbf{m~}%
f\left(  \mathbf{m},N-n\right)  \psi^{\frac{1}{2}\left(  N-n\right)
-1}F^{\left(  n\right)  }\left[  y;\mathbf{m}\right]  ,
\end{equation}
where $F^{\left(  n\right)  }\left[  y;\mathbf{m}\right]  $:
\begin{align}
&  \simeq\exp\left(  -\eta\sum_{k=1}^{n}\epsilon_{k}\right)  \left\{
1+\frac{1}{N}\left[  \frac{1}{2}\eta\left(  \sum_{k=1}^{n}\Delta\mathbf{m}%
_{k}\right)  ^{2}+\right.  \right.  \nonumber\\
&  \left.  \left.  -\eta^{2}\left(  \sum_{k=1}^{n}\epsilon_{k}\right)
^{2}+\left(  n+2\right)  \eta\sum_{k=1}^{n}\epsilon_{k}\right]  \right\}
+O\left(  \frac{1}{N}\right)  ,\label{f1m}%
\end{align}
being $\eta=\psi^{-1}$, the dimensionless inverse temperature. The progressive
calculation demands a refinement of the steepest descend method in order to
account for the $1/N$-contributions obtained beyond of the Gaussian estimation
(see in appendix \ref{refinement}).

\begin{figure}
[t]
\begin{center}
\includegraphics[
height=2.8383in,
width=3.5405in
]%
{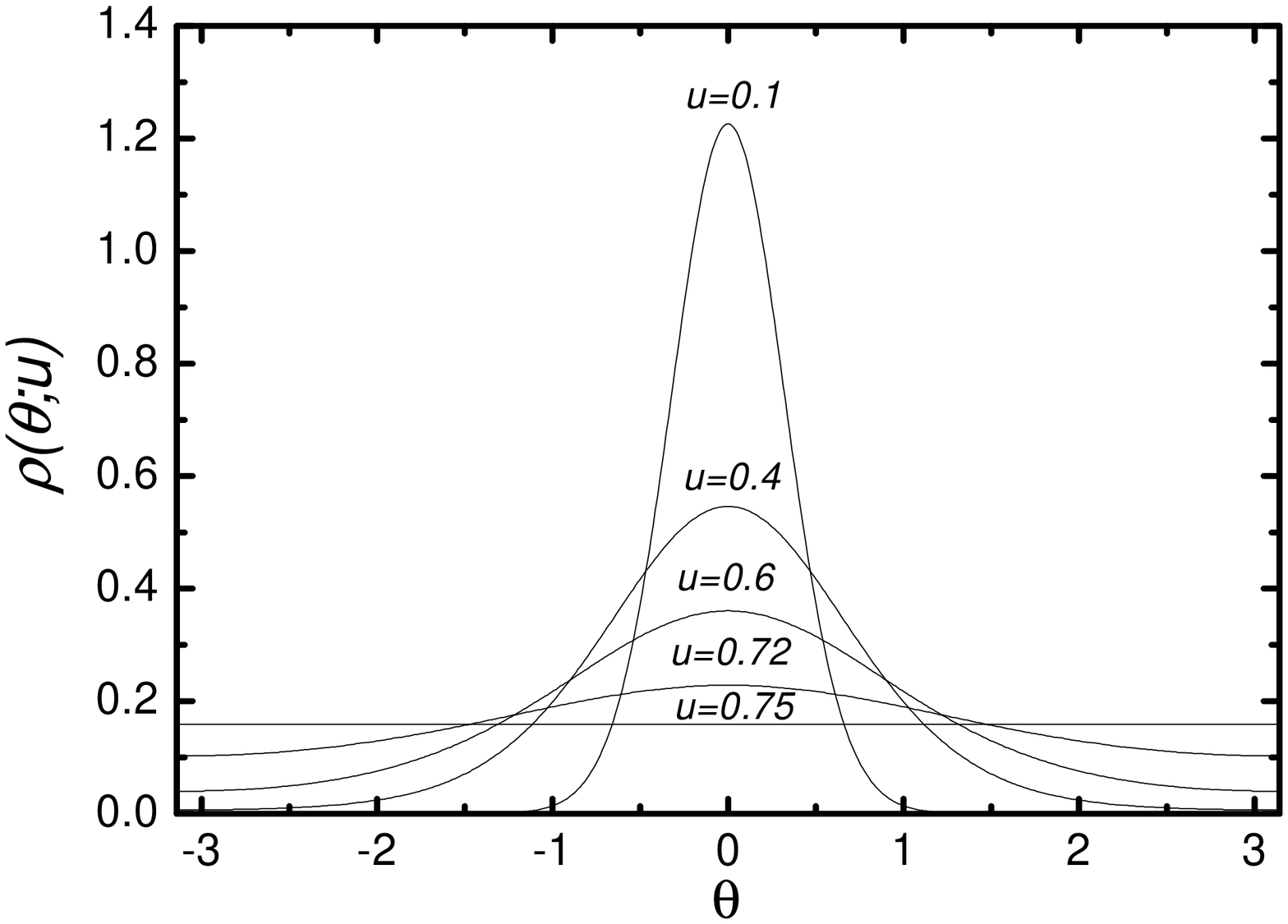}%
\caption{Angular distribution function $\rho\left(  \theta;u\right)  $ for
different energies values. The clustered distributions in the ferromagnetic
phase progressively become in a uniform distribution when the energy $u\geq
u_{c}$. }%
\label{distribution2.eps}%
\end{center}
\end{figure}

The zero-order approximation of the one-body distribution function is\ given by:%

\begin{equation}
f_{m}\left(  \theta,p;u\right)  =C\exp\left(  -\eta\varepsilon\right)  ,
\label{zero-order}%
\end{equation}
where $\varepsilon=\varepsilon\left(  \theta,p\right)  =\frac{1}{2}%
p^{2}-\mathbf{m}\cdot\mathbf{m}\left(  \theta\right)  $, the normalization
constant is given by:%

\begin{equation}
C^{-1}=\sqrt{\frac{2\pi}{\eta}}2\pi I_{0}\left(  \eta m\right)  .
\end{equation}
and the magnetization vector $\mathbf{m}$ satisfied the self-consistent relation:%

\begin{equation}
\mathbf{m}=\int d\theta dp~\mathbf{m}\left(  \theta\right)  f_{m}\left(
\theta,p;u\right)  .
\end{equation}
FIG.\ref{distribution2.eps} shows the angular distribution function\ $\rho
\left(  \theta;u\right)  $ for different energies:%

\begin{equation}
\rho\left(  \theta;u\right)  =\int dp~f_{m}\left(  \theta,p;u\right)
\equiv\frac{\exp\left(  x\cos\theta\right)  }{2\pi I_{0}\left(  x\right)
},\label{den}%
\end{equation}
where the relation $x\equiv\eta m$ was taken into account. Notice that the
ferromagnetic states with $u<u_{c}$ are characterized by the existence of a
clustered distribution of the angular variables $\theta_{k}$ around the
direction of the magnetization vector $\mathbf{m}$, while a uniform
distribution is observed in the paramagnetic phase with $u>u_{c}$. This is the
reason why the ferromagnetic states are referred as a \textit{clustered
phase}, while the paramagnetic ones are referred as a \textit{homogeneous
phase}.%

The terms of the two-body correlation function $g_{m}^{\left(  2\right)
}\left(  x_{1},x_{2}\right)  =F_{m}^{\left(  2\right)  }\left(  x_{1}%
,x_{2}\right)  -F_{m}^{\left(  1\right)  }\left(  x_{1}\right)  F_{m}^{\left(
2\right)  }\left(  x_{2}\right)  $ surviving the first-order approximation are
given by:%

\begin{equation}
g_{m}^{\left(  2\right)  }\left(  x_{1},x_{2}\right)  \simeq\frac{1}{N}%
f_{1}f_{2}\left[  \Phi_{12}+\eta^{2}g\left(  u\right)  \Theta_{12}\right]
+O\left(  \frac{1}{N}\right)  ,\ \label{gmicro}%
\end{equation}
where $f_{i}=f_{m}\left(  \theta_{i},p_{i};u\right)  $ is the zero-order
approximation of the one-body distribution function (\ref{zero-order}),
$g\left(  u\right)  =N\left\langle \Delta\mathbf{m}^{2}\right\rangle $, the
correlation function derived\ from the average magnetization dispersion
$\left\langle \Delta\mathbf{m}^{2}\right\rangle $ showed in
FIG.\ref{micro.eps}, and $\Phi_{12}$ and $\Theta_{12}$, two new form factors
given by:%
\begin{align}
\Phi_{12} &  =\eta\delta\mathbf{m}_{1}\cdot\delta\mathbf{m}_{2}-2\eta
^{2}\delta\varepsilon_{1}\delta\varepsilon_{2},\\
\Theta_{12} &  =\left(  2\eta\mathbf{m}\delta\varepsilon_{1}+\delta
\mathbf{m}_{1}\right)  \cdot\left(  2\eta\mathbf{m}\delta\varepsilon
_{2}+\delta\mathbf{m}_{2}\right)  .
\end{align}
The quantities $\delta\varepsilon_{i}=\varepsilon_{i}-\left\langle
\varepsilon\right\rangle $ and $\delta\mathbf{m}_{i}=\mathbf{m}_{i}%
-\mathbf{m}$ represent the energy and magnetization desviations respectively.
The presence of the function $g\left(  u\right)  $ in Eq.(\ref{gmicro}) leads
to the divergence of the two-body correlation function at the critical point
$u_{c}$. The spacial two-body correlation function in the homogeneous phase is
given by $g\left(  \theta_{1},\theta_{2}\right)  =\left(  2\pi\right)
^{-2}c_{2}\left(  \theta_{1},\theta_{2};u\right)  $, where the function:%
\begin{equation}
c\left(  \theta_{1},\theta_{2}\right)  =\left(  1+\frac{1}{4}\frac{1}{u-u_{c}%
}\right)  \frac{1}{2u-1}\cos\left(  \theta_{1}-\theta_{2}\right)  ,\label{cc}%
\end{equation}
diverges at the critical point in terms of the inverse temperature $\eta$ as
follows $\simeq4\left(  \eta_{c}-\eta\right)  ^{-1}\cos\left(  \theta
_{1}-\theta_{2}\right)  $. This asymptotic behavior is consistent with the one
estimated in the refs.\cite{chava,chava2,chava3}, but (\ref{cc}) is now the
exact microcanonical result within the first-order approximation.

The inexistence of three-body terms in Eq.(\ref{f1m}) straightforwardly leads
to the vanishing of the three-body correlation function $g_{m}^{\left(
3\right)  }\left(  x_{1},x_{2},x_{3}\right)  $:%

\begin{align}
&
\begin{tabular}
[c]{c}%
$g_{m}^{\left(  3\right)  }\left(  x_{1},x_{2},x_{3}\right)  =F_{m}^{\left(
3\right)  }\left(  x_{1},x_{2},x_{3}\right)  -g_{m}^{\left(  2\right)
}\left(  x_{1},x_{2}\right)  F_{m}^{\left(  1\right)  }\left(  x_{3}\right)  $%
\end{tabular}
\nonumber\\
&
\begin{tabular}
[c]{c}%
$-g_{m}^{\left(  2\right)  }\left(  x_{2},x_{3}\right)  F_{m}^{\left(
1\right)  }\left(  x_{1}\right)  -g_{m}^{\left(  2\right)  }\left(
x_{3},x_{1}\right)  F_{m}^{\left(  1\right)  }\left(  x_{2}\right)  ,$%
\end{tabular}
\end{align}
in the first-order approximation, and hence, the three-body correlations are
just $O\left(  1/N\right)  $ size effects. Such a microcanonical result will
be taken into consideration during the derivation of suitable dynamical
equations for the distribution and correlation functions based on the
well-known \textit{BBGKY hierarchy}.

\section{Thermodynamic stability}

Thermodynamic stability concerns to the question about when a given admissible
macrostate characterized by a certain energy and magnetization is stable or
unstable under thermal fluctuations associated to the thermodynamic
equilibrium where the temperature and the magnetic field act as constant
control parameters. Such a question is also intimately related to the nature
of the correspondence between the control parameters $\left(  \eta,H\right)  $
and the controlled system observables $\left(  u,m\right)  $ (internal energy
$u$ and the projection of the magnetization $m$ along the magnetic field),
that is, the existence or inexistence of the \textit{ensemble equivalence}.

The analysis starts from the consideration of the partition function $Z=\int
dX_{N}\exp\left[  -\beta H_{N}^{\ast}\right]  $ and the introduction of the
Planck thermodynamical potential per particle $p\left(  \eta,H\right)  =-\ln
Z\left(  \eta,H\right)  /N$, which can be rephrased as follows:
\begin{equation}
\exp\left[  -Np\left(  \eta,H\right)  \right]  \propto\int dmdu\exp\left\{
-Np\left(  \eta,H;u,m\right)  \right\}  ,\label{planck}%
\end{equation}
where $p\left(  \eta,H;u,m\right)  =\eta u+\lambda m-s\left(  u,m\right)  $
with $\lambda=-\beta H$ and the \textit{detailed entropy} $s\left(
u,m\right)  $ given in a parametric form by:
\begin{equation}
s\left(  u,m\right)  =s_{0}+\frac{1}{2}\ln\left\{  \kappa+m^{2}\left(
x\right)  \right\}  -xm\left(  x\right)  +\ln J_{0}\left(  x\right)
\label{detail s}%
\end{equation}
within the approximation provided by the steepest descend method. The
exponential function of the integral (\ref{planck}) exhibits sharp peaks
around its maxima when $N$ is large enough, a behavior that allows to rephrase
the integral (\ref{planck}) within the Gaussian approximation as follows:%
\begin{equation}
\sim\sum_{k}\exp\left(  -Np_{k}\right)  \int\exp\left[  \frac{1}{2}NB\left(
u,m;u_{k},m_{k}\right)  \right]  dmdu,
\end{equation}
where $p_{k}=p\left(  \eta,H;u_{k},m_{k}\right)  $ and the bilinear form
$B\left(  u,m;u_{k},m_{k}\right)  =H_{uu}\Delta u_{k}^{2}+\left(
H_{um}+H_{mu}\right)  \Delta u_{k}\Delta m_{k}+H_{mm}\Delta m_{k}^{2}$ with
$\Delta u_{k}=u-u_{k}$ and $\Delta m_{k}=m-m_{k}$ obtained from the entropy
Hessian $H_{ij}$:
\begin{equation}
H_{ij}=\left(
\begin{array}
[c]{cc}%
H_{uu} & H_{mu}\\
H_{um} & H_{mm}%
\end{array}
\right)  =\left(
\begin{array}
[c]{cc}%
\partial^{2}s/\partial u^{2} & \partial^{2}s/\partial m\partial u\\
\partial^{2}s/\partial u\partial m & \partial^{2}s/\partial m^{2}%
\end{array}
\right)  .\label{hessian}%
\end{equation}
evaluated at the \textit{k}-th stable stationary point:
\begin{equation}
\beta=\frac{\partial s\left(  u_{k},m_{k}\right)  }{\partial u},~\lambda
=\frac{\partial s\left(  u_{k},m_{k}\right)  }{\partial m},\label{stationary}%
\end{equation}
which satisfies the negative definition of the Hessian matrix (\ref{hessian}).
From the thermodynamical point of view, such maxima represent the coexisting
macrostates or phases appearing for given values of $\beta$ and $H$.
Obviously, there is ensemble inequivalence when there exist only one maximum
representing a unique stable phase with given values of $\left(  u,m\right)
$. The greatest peak during the phase coexistence corresponds to the stable
phase, while the other represent metastable states. The greatest peak
corresponding to the stable phase provides the main contribution of the
integral (\ref{planck}) allowing to estimate the Planck thermodynamic
potential per particle as follows:
\begin{equation}
p\left(  \eta,H\right)  =\inf\left\{  \eta u+\lambda m-s\left(  u,m\right)
\right\}  .
\end{equation}
This is\ just the Legendre transformation which constitutes a fundamental
stone of the thermodynamic formalism \cite{gallavotti}. All those admissible
macrostates of the system satisfying the stationary condition
(\ref{stationary}) but do not obey the negative definition of the entropy
Hessian matrix (\ref{hessian}) are precisely the unstable macrostates.%

\begin{figure}
[t]
\begin{center}
\includegraphics[
height=2.6238in,
width=3.5405in
]%
{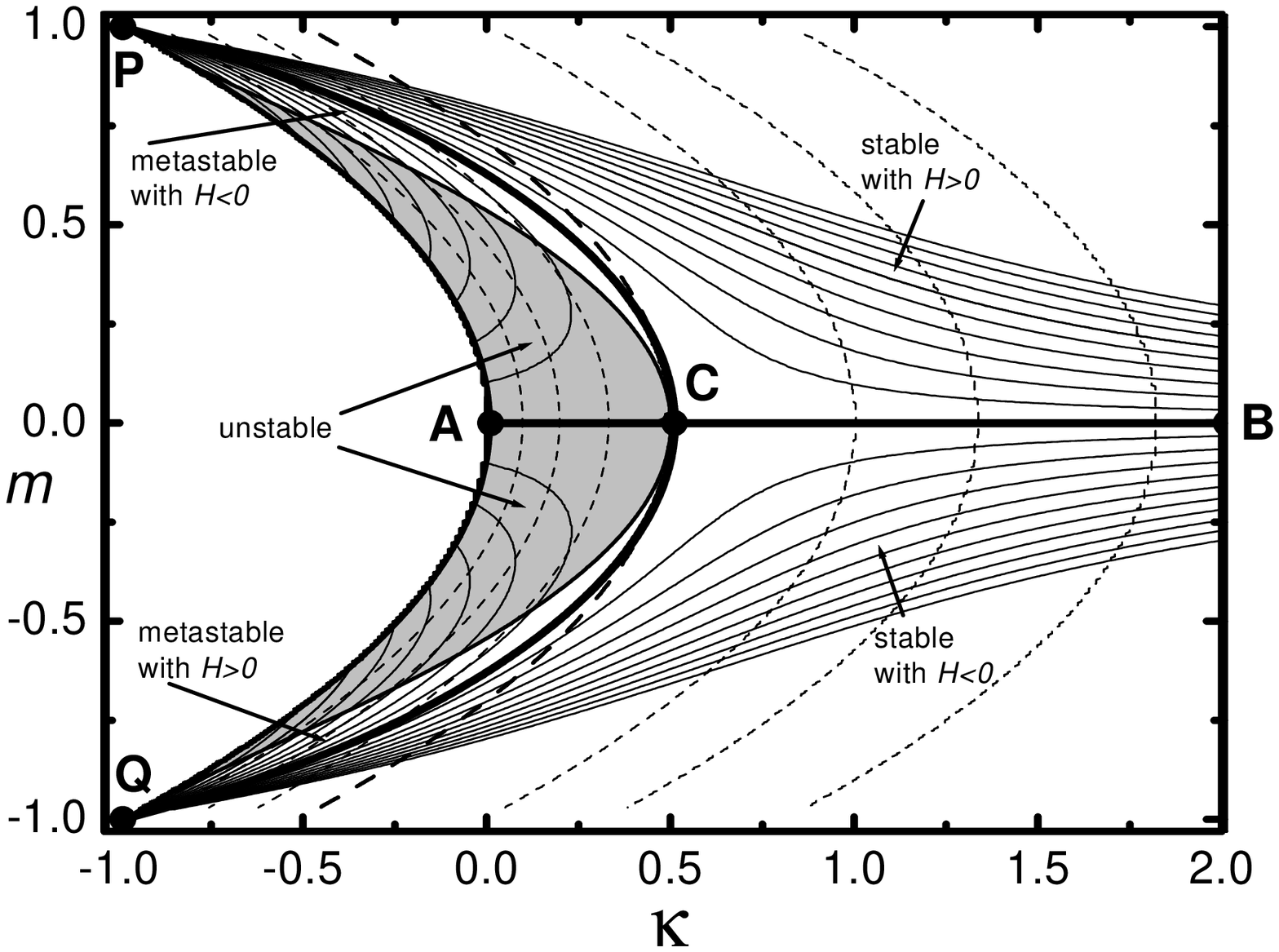}%
\caption{Magnetization curves of the HMF model and their stability in the
plane $\left(  \kappa,m\right)  $. Solid lines: magnetization curves at
constant magnetic field $H\not =0$; Thick solid lines: magnetization curves at
$H=0$; Dashed lines: isothermal magnetization curves. Zones outside the
curvilinear region \textbf{PCQA} are canonically stable. The dark zone inside
this region is canonically unstable, that is, the region of ensemble
inequivalence, while the white zones enclose the \textit{metastable states}.
Point \textbf{C}: critical point of the continuous phase transition.}%
\label{unstable.magnet.eps}%
\end{center}
\end{figure}
%

\begin{figure}
[tb]
\begin{center}
\includegraphics[
height=2.636in,
width=3.5405in
]%
{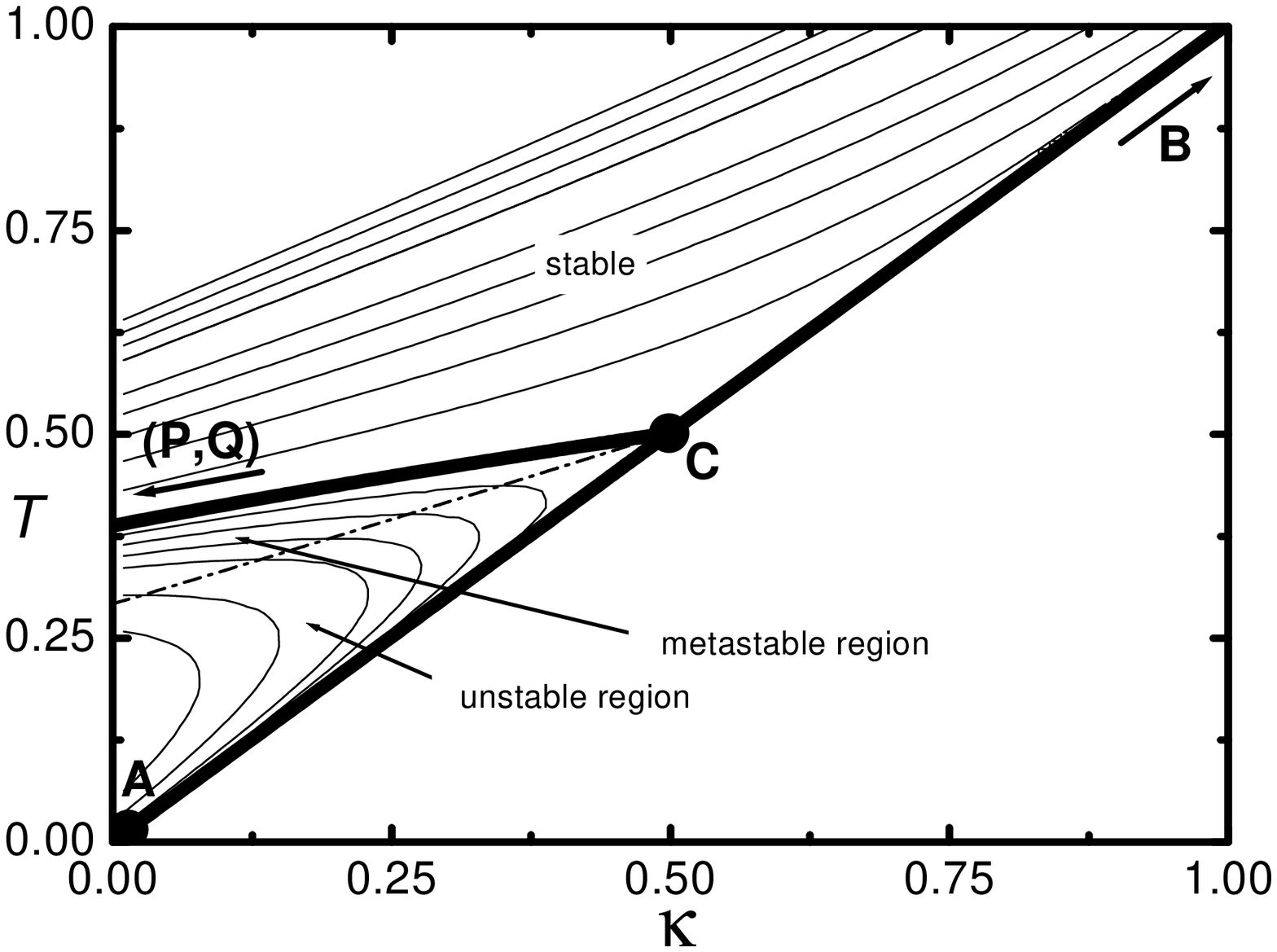}%
\caption{Caloric curves of the HMF model in the neighborhood of the critical
point \textbf{C}. Solid lines: caloric curves at constant magnetic field
$H\not =0$. Thick solid lines: caloric curves at $H=0$, where \textbf{BC}%
-(\textbf{PQ}) and \textbf{CA} are the stable and unstable branches
respectively. The nonanalyticity of the canonical thermodynamic potential at
the critical point C follows from the bifurcation undergone by the caloric
curve at $H=0$ in this point. The dash-dotted line divided the caloric curves
into metastable and unstable regions.}%
\label{unstable.caloric.eps}%
\end{center}
\end{figure}

Performing the calculations, the canonical parameters:%

\begin{equation}
\beta=\frac{\partial s}{\partial u}=\frac{1}{\kappa+m^{2}\left(  x\right)
},~\lambda=\frac{\partial s}{\partial m}=\frac{m}{\kappa+m^{2}\left(
x\right)  }-x,
\end{equation}
allow to express the magnetic field $H$ and the temperature $T=\beta^{-1}$ in
terms of the macroscopic observables $\kappa$ and $m$:
\begin{align}
H &  =H\left(  \kappa,m\right)  =x\left(  \kappa+m^{2}\left(  x\right)
\right)  -m\left(  x\right)  ,\nonumber\\
T &  =T\left(  \kappa,m\right)  =\kappa+m^{2}\left(  x\right)  .
\end{align}
According to the entropy per particle (\ref{detail s}), the physically
admissible regions satisfy the inequalities $\kappa+m^{2}>0$ and $m^{2}\leq1$.
The regions of stability in the plane $\left(  \kappa,m\right)  $ are
determined from the entropy Hessian:%
\begin{equation}
H_{ij}=\beta^{2}\left(
\begin{array}
[c]{cc}%
-2 & -2m\\
-2m & \kappa-m^{2}-a\left(  \kappa+m^{2}\right)  ^{2}%
\end{array}
\right)  ,\nonumber
\end{equation}
being $a^{-1}=m^{\prime}\left(  x\right)  \equiv1-m\left(  x\right)
/x-m^{2}\left(  x\right)  $. The determinant $D=\det H_{ij}=-2\left(
\kappa+m^{2}\right)  ^{-3}\left[  1-a\left(  \kappa+m^{2}\right)  \right]  $
vanishes at the boundary of the unstable region, a curve which is
parametrically represented as:%
\begin{equation}
\left[  \kappa\left(  x\right)  ,m\left(  x\right)  \right]  =\left[
1-\frac{m\left(  x\right)  }{x}-2m^{2}\left(  x\right)  ,m\left(  x\right)
\right]  .
\end{equation}

These results are summarized in FIG.\ref{unstable.magnet.eps} and
FIG.\ref{unstable.caloric.eps}. The curve \textbf{PAQ} is the boundary of the
physically admissible macrostates. The thick solid lines represent the
magnetization curves at $H=0$. The stable branch \textbf{PCQ} constitutes the
boundary between the stable and metastable regions. The critical point of the
continuous phase transitions \textbf{C}$:\left(  \kappa_{c},0\right)  $\ with
$\kappa_{c}=2u_{c}-1=0.5$ is just a \textit{bifurcation point} of the
magnetization curves located at the endpoint of the metastable region which
touches also the stable and the unstable regions, and therefore, it is a point
of marginal stability.

The unstable region (dark zones in the plane $\left(  \kappa,m\right)  $ shown
in FIG.\ref{unstable.magnet.eps}) is characterized by the presence of negative
values of the magnetic susceptibility at constant temperature $\chi
_{H}^{\left(  c\right)  }=\left(  \partial M/\partial H\right)  _{T}$, a
behavior canonically anomalous for this kind of model system in terms of the
well-known thermodynamical identity $\chi_{H}^{\left(  c\right)  }%
=\beta\left\langle \delta M^{2}\right\rangle $, but which is microcanonically
\textit{admissible}. Actually, it possesses the same anomalous character of
the macrostates with a negative heat capacity $C=dE/dT<0$ observed in several
systems \cite{pad}\ in terms of the thermodynamical identity $C=\beta
^{2}\left\langle \delta E^{2}\right\rangle $.

The lost of analyticity in the thermodynamic limit of the thermodynamic
potentials like the microcanonical entropy per particle $s\left(
\varepsilon,H\right)  =\sup_{m}\left\{  s\left[  u\left(  =\varepsilon
+Hm\right)  ,m\right]  \right\}  $ or the Helmholtz Free energy per particle
$f\left(  \beta,H\right)  =\inf_{\varepsilon}\left\{  \varepsilon-Ts\left(
\varepsilon,H\right)  \right\}  \equiv\inf_{u,m}\left\{  u-Hm-Ts\left(
u,m\right)  \right\}  $ at the critical point \textbf{C} can be related to the
bifurcation of the magnetization curve shown in FIG.\ref{unstable.magnet.eps}
or caloric curve show in FIG.\ref{unstable.caloric.eps} for $H=0$. The system
is unable to follows the "trajectory" \textbf{BCA} with $m=0$ since the branch
\textbf{CA} is located inside the region of ensemble inequivalence where there
exist anomalous macrostates with $\chi_{H}^{\left(  c\right)  }<0$. The large
thermodynamic fluctuations existing there provoke a sudden change of the
original tendency \textbf{BCA} following in this way anyone of the stable
symmetric branches \textbf{CP} or \textbf{CQ} with a nonvanishing
magnetization $m\not =0$, and consequently, the occurrence of a
\textit{spontaneous symmetry breaking} \cite{gallavotti}.

\section{Some final remarks}

As already illustrated, the thermodynamic properties of the HMF model do not
essentially differ from the other ferromagnetic models with short-range
interactions. We have previously shown that the relevant microcanonical
thermodynamic variable is $u=E/gN^{2}$, and the characteristic energy is
$E_{0}=gN^{2}$. The thermodynamic limit is carried out when\ $N$ is sent to
the infinity by keeping fixed the dimensionless energy $u$. This is the same
thermodynamic limit introduced in ref.\cite{chava} in order to perform the
mean field description of this model. However, our analysis reveals that the
entropy per particles is ill-defined in the thermodynamic limit: while the
term of $s\left(  u,N;I,g\right)  $ containing the relevant thermodynamical
variable $u$ is \textit{N}-independent, the \textit{N-}dependent additive
constant $s_{0}=%
\frac12
\ln\left(  2\pi e^{2}Ig/N\right)  $ \textit{diverges} when $N\rightarrow
\infty$ (see in ref.\cite{chava4}). It means that the ill-behavior of the
entropy per particle is \textit{unavoidable} without considering an
appropriate \textit{N}-dependence for the coupling constant $g\,$.

The using of an appropriate \textit{N}-dependence in the coupling constant $g$
is usually identified with the \textit{Kac prescription} \cite{kac}. The
standard usage of this procedure is to consider certain dependence $g\left(
N\right)  $ that ensures the \textit{extensive growing} of the total energy
$E$, since the energy per particle $e=$ $E/N$ is keep fixed in the
thermodynamic limit $N\rightarrow\infty$. This condition demands that
$E_{0}/N=gN=\gamma=const$. Although most of works devoted to the HMF model
make use of this condition
\cite{ant,lat1,lat2,lat3,lat4,zanette,dauxois,yamaguchi,ybd,chava,bouchet1,bouchet2}%
, it is very easy to verify that its application does not avoid the divergence
of the entropy per particle in the thermodynamic limit, $\lim_{N\rightarrow
\infty}s_{0}=\infty$.

A closer look to this question clarifies that this procedure should be applied
with care. Firstly, the coupling constants determine the characteristic
temporal scales acting on the dynamical evolution of a given system, and
therefore, any scaling \textit{N}-dependence of these coupling constant
affects the system dynamical behavior during the imposition of the
thermodynamic limit $N\rightarrow\infty$. This obvious remark is very
important to take into consideration since the results of many numerical
simulations evidence the \textit{noncommutativity} of the thermodynamic limit
$N\rightarrow\infty$ with the infinite time limit $T\rightarrow\infty$
necessary for the equilibration of temporal averages
\cite{lat2,lat3,lat4,zanette}:
\begin{equation}
\lim_{N\rightarrow\infty}\lim_{T\rightarrow\infty}\left\langle A\right\rangle
\left(  T,N\right)  \not =\lim_{T\rightarrow\infty}\lim_{N\rightarrow\infty
}\left\langle A\right\rangle \left(  T,N\right)  ,
\end{equation}
(where $\left\langle A\right\rangle \left(  T,N\right)  =\int_{0}^{T}A\left[
X_{N}\left(  t\right)  \right]  dt/T$) as a consequence of the divergence of
the relaxation time $\tau_{eq}$ in the thermodynamic limit, $\lim
_{N\rightarrow\infty}\tau_{eq}=\infty$. In the authors opinion, the origin of
this anomaly could be related to an inappropriate use of the Kac prescription.
This argument follows from questioning the main motivation of introducing the
Kac prescription: to deal with an extensive energy. The extensivity of the
energy is a thermodynamic feature of the extensive systems intimately related
to the statistical independence or separability of a large system in
independent subsystems appearing as a consequence of the incidence of
short-range forces. Obviously, such a microscopic picture is outside the
context within the HMF model, where the long-range character of the
microscopic interactions implies its \textit{intrinsic nonextensive nature}.

The HMF model could be consider as a limit case with $\alpha\rightarrow0$ of a
parametric family of ferromagnetic models on a square lattice \cite{celia}
whose potential energy is given by:
\begin{equation}
V_{\alpha}=\frac{1}{2}g\sum_{i=1}^{N}\sum_{j=1}^{N}\frac{1}{r_{ij}^{\alpha}%
}\left[  1-\cos\left(  \theta_{i}-\theta_{j}\right)  \right]  ,
\end{equation}
where $r_{ij}$ is the lattice distance between the i-th and j-th rotator. The
integral estimation of potential energy per particle in the paramagnetic phase
(where the averages $\left\langle \cos\left(  \theta_{i}-\theta_{j}\right)
\right\rangle \approx0$ ) by using a lattice with spacing $a$ and linear
dimension $R$ $\propto\sqrt{N}$:%
\begin{align}
\upsilon_{\alpha} &  =V_{\alpha}/N\sim a^{-2}\int_{a}^{R}\frac{2\pi
rdr}{r^{\alpha}}\nonumber\\
&  =2\pi\left\{
\begin{array}
[c]{cc}%
\left(  \alpha-2\right)  ^{-1}\left(  a^{2-\alpha}-R^{2-\alpha}\right)   &
\alpha\not =2\\
\ln\left(  R/a\right)   & \alpha=2
\end{array}
\right.
\end{align}
allows to understand that such as models are extensive in the thermodynamic
limit when $\alpha>2$ and nonextensive elsewhere. As a consequence of the
existence of long-range correlations, the nonextensive system cannot
be\ trivially divided in independent subsystems, and therefore, there is no
physical reason to justify the imposition of the extensive character of the
total energy in this case.

The using a suitable scaling dependence for the coupling constants in order to
regularize the thermodynamic parameters and potentials of a given system in
the thermodynamic limit might be applicable in some way that also ensures a
well-defined dynamical behavior characterized by the commutativity of the
limits $\lim_{N\rightarrow\infty}$ and $\lim_{T\rightarrow\infty}$. Generally
speaking, all that it is necessary to demand is the non divergence of the
relaxation timescale $\tau_{eq}$ when $N\rightarrow\infty$. With some recent
exceptions \cite{yamaguchi,ybd}, most of numerical studies of the microscopic
dynamics of the HMF model revealed a characteristic relaxation timescale for
the HMF model given by $\tau_{eq}=\tau_{mic}N$, being $\tau_{mic}=\sqrt{I/Ng}$
the characteristic microscopic time of the rotators
evolution\ \cite{lat1,lat2,lat3,lat4,zanette,dauxois}. It is easy to realize
that the imposition of the scaling $g/N=\gamma=const$ ensures both the
well-defined behavior of the additive constant of the entropy per particle
$s_{0}=%
\frac12
\ln\left(  2\pi e^{2}Ig/N\right)  $ and the relaxation timescale $\tau
_{eq}=\sqrt{IN/g}$ when $N\rightarrow\infty$. This ansatz leads to a power-low
growing of the characteristic energy $E_{0}\propto N^{3}$, which does not
involves in principle any physical inconsistency since the HMF model is
nonextensive. We shall return to this discussion in the forthcoming papers.

\appendix

\section{Demonstrations}

\subsection{Derivation of Eq.(\ref{func})\label{expan}}

We perform the power expansion up to the second order approximation of
$\mathbf{k}$ for the complex number $z$:%

\begin{equation}
z=-i\sqrt{\mathbf{K}^{2}}\simeq x+i\mathbf{n\cdot k-}\frac{1}{2x}%
\mathbf{k}^{2}-\frac{1}{2x}\left(  \mathbf{n\cdot k}\right)  ^{2}+O\left(
\left\vert \mathbf{k}\right\vert ^{2}\right)  ,
\end{equation}
being $x=\left\vert \mathbf{x}\right\vert $ and $\mathbf{n}=\mathbf{x}/x$. Let
us now carry out the power expansion for the function $\ln I_{0}\left(
x+\Delta z\right)  $ in term of the variable $\Delta z$ up to the second order
approximation, being $\Delta z=z-x$:%

\begin{equation}
\ln I_{0}\left(  x+\Delta z\right)  =+m\left(  x\right)  \Delta z+\frac{1}%
{2}\frac{dm\left(  x\right)  }{dx}\left(  \Delta z\right)  ^{2}+O\left[
\left(  \Delta z\right)  ^{2}\right]  .
\end{equation}
where $m\left(  x\right)  =I_{1}\left(  x\right)  /I_{0}\left(  x\right)  $.
Thus, the exponential function:%

\begin{equation}
\exp\left[  N\mathbf{K}\cdot\mathbf{m}+\ln I_{0}\left(  z\right)  \right]  ,
\end{equation}
can be rewritten by dismissing the terms $O\left(  \left\vert \mathbf{k}%
\right\vert ^{2}\right)  $ as follows:%

\begin{align}
&  \simeq\exp\left\langle -N\left[  xm-\ln I_{0}\left(  x\right)  \right]
\right\rangle \times\nonumber\\
&  \times\exp\left\langle -\frac{N}{2}\left[  \kappa_{1}\left(  x\right)
k_{1}^{2}-\kappa_{2}\left(  x\right)  k_{2}^{2}\right]  \right\rangle
,\label{expk}%
\end{align}
where $\mathbf{x\parallel m}$ and $m=\left\vert \mathbf{m}\right\vert $;
$k_{1}^{2}=\left(  \mathbf{n}\cdot\mathbf{k}\right)  ^{2}$ and $k_{2}%
^{2}=\mathbf{k}^{2}-\left(  \mathbf{n}\cdot\mathbf{k}\right)  ^{2}$; while
$\kappa_{1}\left(  x\right)  =2m\left(  x\right)  /x+dm\left(  x\right)  /dx$
and $\kappa_{2}\left(  x\right)  =m\left(  x\right)  /x$. Since $m\left(
x\right)  $ is an odd monotonic increasing function, with $m\left(  x\right)
\geq0$ when $x\geq0$, the functions $\kappa_{1}\left(  x\right)  $ and
$\kappa_{2}\left(  x\right)  $ are always nonnegative. The integration of the
expression (\ref{expk}) leads to the result shown in Eq.(\ref{func}).

\subsection{Refinement of the steepest descend method\label{refinement}}

The calculation of the first-order approximation of the correlation functions
involves a little refinement of the steepest descend method. The aim is the
determination of the $1/N$-contributions of the average of a physical quantity
$A\left(  \mathbf{x}\right)  $ as follows:
\begin{equation}
\left\langle A\right\rangle \Omega=\int A\left(  \mathbf{x}\right)
\exp\left[  Ns\left(  \mathbf{x};a\right)  \right]  d^{n}\mathbf{x,}
\label{a11}%
\end{equation}
where the partition function $\Omega$ is given by:%
\begin{equation}
\Omega=\int\exp\left[  Ns\left(  \mathbf{x};a\right)  \right]  d^{n}%
\mathbf{x.}%
\end{equation}

Firstly, we perform the power expansion of the function $s\left(
\mathbf{x}\right)  $ around its maximum point $\mathbf{x}_{0}$ as follows:%
\begin{equation}
s\left(  \mathbf{x}\right)  \simeq s\left(  \mathbf{x}_{0}\right)  +c_{2}%
\cdot\left(  \Delta\mathbf{x}\right)  ^{2}+R\left(  \Delta\mathbf{x}%
;\mathbf{x}_{0}\right)  , \label{power1}%
\end{equation}
where the higher-order contributions $R\left(  \Delta\mathbf{x};\mathbf{x}%
_{0}\right)  $ are denoted by:
\begin{equation}
R\left(  \Delta\mathbf{x};\mathbf{x}_{0}\right)  =\sum_{m=3}^{\infty}%
c_{m}\cdot\left(  \Delta\mathbf{x}\right)  ^{m}.
\end{equation}
Hereafter, we shall consider the following convention in order to simplify the
notation during the calculation. For example, the term $c_{m}\cdot\left(
\Delta\mathbf{x}\right)  ^{m}$ represents the tensorial product:
\begin{equation}
c_{m}\cdot\left(  \Delta\mathbf{x}\right)  ^{m}\equiv\frac{1}{m!}%
\sum_{\left\{  i\right\}  }\frac{\partial^{m}s\left(  \mathbf{x}_{0}\right)
}{\partial x^{i_{1}}\partial x^{i_{2}}\ldots\partial x^{i_{m}}}\Delta
x^{i_{1}}\Delta x^{i_{2}}\ldots\Delta x^{i_{m}}.
\end{equation}

Eq.(\ref{power1}) allows to express the exponential function as follows:
\begin{equation}
\exp\left[  Ns\left(  \mathbf{x}\right)  \right]  =\exp\left[  Ns\left(
\mathbf{x}_{0}\right)  \right]  \exp\left[  -\frac{1}{2}\omega\cdot\left(
\Delta\mathbf{x}\right)  ^{2}\right]  F\left(  \Delta\mathbf{x};\mathbf{x}%
_{0}\right)  ,
\end{equation}
where $\omega=-2Nc_{2}$ and the function $F\left(  \Delta\mathbf{x}%
;\mathbf{x}_{0}\right)  =\exp\left[  NR\left(  \Delta\mathbf{x};\mathbf{x}%
_{0}\right)  \right]  $ is given by:%
\begin{equation}
F\left(  \Delta\mathbf{x};\mathbf{x}_{0}\right)  =\sum_{k=0}^{\infty}\frac
{1}{k!}N^{k}R^{k}\left(  \Delta\mathbf{x};\mathbf{x}_{0}\right)  .
\end{equation}
This latter result can be conveniently rewritten as a power series expansion:%
\begin{equation}
F\left(  \Delta\mathbf{x};\mathbf{x}_{0}\right)  =\sum_{m=0}^{\infty}%
\kappa_{m}\cdot\left(  \Delta\mathbf{x}\right)  ^{m},\label{power2}%
\end{equation}
where%
\begin{equation}
\kappa_{m}=\sum_{\left\{  j\right\}  }\delta\left(  m-\sum_{k=3}^{\infty
}kj_{k}\right)  \prod_{k=3}^{\infty}\frac{N^{j_{k}}c_{k}^{j_{k}}}{j_{k}!},
\end{equation}
being $\delta\left(  n\right)  $ an integer function defined by:%
\begin{equation}
\delta\left(  n\right)  =\left\{
\begin{array}
[c]{cc}%
1 & n=0\\
0 & n\not =0
\end{array}
\right.  .
\end{equation}
It can be checked that the first 10 coefficients of the power expansion
(\ref{power2}) are given by:%
\begin{align}
\kappa_{0} &  =1,~\kappa_{1}=\kappa_{2}=0,~\kappa_{3}=Nc_{3},~\kappa
_{4}=Nc_{4},\nonumber\\
~\kappa_{5} &  =Nc_{5},~\kappa_{6}=Nc_{6}+\frac{1}{2!}N^{2}c_{3}^{2}%
,~\kappa_{7}=Nc_{7}+N^{2}c_{3}c_{4},~\nonumber\\
\kappa_{8} &  =Nc_{8}+N^{2}\left(  c_{3}c_{5}+\frac{1}{2!}c_{4}^{2}\right)
,~\\
\kappa_{9} &  =Nc_{9}+N^{2}\left(  c_{3}c_{6}+c_{4}c_{5}\right)  +\frac{1}%
{3!}N^{3}c_{3}^{3},\nonumber\\
\kappa_{10} &  =Nc_{10}+N^{2}\left(  c_{3}c_{7}+c_{4}c_{6}+\frac{1}{2!}%
c_{5}c_{5}\right)  +\frac{1}{2!}N^{3}c_{3}^{2}c_{4}.\nonumber
\end{align}

Taking into account that in the Gaussian integration:
\begin{equation}
\left\langle \left(  \Delta\mathbf{x}\right)  ^{n}\right\rangle =N_{0}\int
d^{n}\mathbf{x}\exp\left[  -\frac{1}{2}\omega\cdot\left(  \Delta\mathbf{x}%
^{2}\right)  \right]  \left(  \Delta\mathbf{x}\right)  ^{k},
\end{equation}
survive only the odd dispersions, denoting by $g_{2k}=N^{k}\left\langle
\left(  \Delta\mathbf{x}\right)  ^{2k}\right\rangle $, being $N_{0}=\sqrt
{\det\left(  \frac{1}{2\pi}\omega\right)  }$ the normalization constant, the
first-order approximation of the partition function $\Omega$ is given by:%
\begin{equation}
\Omega=\Omega_{0}\left\{  1+\frac{1}{N}\left(  c_{4}g_{4}+\frac{1}{2}c_{3}%
^{2}g_{6}\right)  \right\}  ,
\end{equation}
where $\Omega_{0}=\exp\left[  Ns\left(  \mathbf{x}_{0}\right)  \right]
/N_{0}$ is the zero-order approximation of the partition function.

The calculation of the integral (\ref{a11}) is carried out by performing the
power expansion of the quantity $A\left(  \mathbf{x}\right)  $ around the
maximum point $\mathbf{x}_{0}$%
\begin{equation}
A\left(  \mathbf{x}\right)  =\sum_{m=0}^{\infty}A^{\left(  m\right)  }%
\cdot\left(  \Delta\mathbf{x}\right)  ^{m},
\end{equation}
being:
\begin{equation}
A^{\left(  m\right)  }\cdot\left(  \Delta\mathbf{x}\right)  ^{m}=\frac{1}%
{m!}\sum_{\left\{  i\right\}  }\frac{\partial^{m}A\left(  \mathbf{x}%
_{0}\right)  }{\partial x^{i_{1}}\partial x^{i_{2}}\ldots\partial x^{i_{m}}%
}\Delta x^{i_{1}}\Delta x^{i_{2}}\ldots\Delta x^{i_{n}},
\end{equation}
which allows to rewrite:
\begin{equation}
A\left(  \mathbf{x}\right)  F\left(  \Delta\mathbf{x};\mathbf{x}_{0}\right)
=\sum_{n}B_{n}\left(  \Delta\mathbf{x}\right)  ^{n},
\end{equation}
being:%
\begin{equation}
B_{n}=\sum_{m=1}^{n}A^{\left(  m\right)  }\kappa_{n-m}.
\end{equation}

Thus, the first-order approximation of the integral (\ref{a11}) is given by:%

\begin{align}
\left\langle A\right\rangle \Omega &  =\Omega_{0}\left\{  A^{\left(  0\right)
}+\frac{1}{N}\left(  A^{\left(  2\right)  }g_{2}+\right.  \right.
,\nonumber\\
&  \left.  \left.  +\left(  A^{\left(  0\right)  }c_{4}+A^{\left(  1\right)
}c_{3}\right)  g_{4}+\frac{1}{2}A^{\left(  0\right)  }c_{3}^{2}\right)
\right\}  ,
\end{align}
which leads to the following result%
\begin{equation}
\left\langle A\right\rangle =A^{\left(  0\right)  }+\frac{1}{N}\left(
A^{\left(  2\right)  }g_{2}+A^{\left(  1\right)  }c_{3}g_{4}\right)  .
\end{equation}
Restoring now the ordinary notation, we finally obtain:
\begin{equation}
\left\langle A\right\rangle =A\left(  \mathbf{x}_{0}\right)  +\frac{1}{2!}%
\sum_{i_{1}i_{2}}\frac{\partial^{2}A\left(  \mathbf{x}_{0}\right)  }{\partial
x^{i_{1}}\partial x^{i_{2}}}\left\langle \Delta x^{i_{1}}\Delta x^{i_{2}%
}\right\rangle +
\end{equation}%
\[
+\frac{1}{3!}\sum_{i_{1}i_{2}i_{3}i_{4}}\frac{\partial A\left(  \mathbf{x}%
_{0}\right)  }{\partial x^{i_{1}}}\frac{\partial^{3}s\left(  \mathbf{x}%
_{0}\right)  }{\partial x^{i_{2}}\partial x^{i_{3}}\partial x^{i_{4}}%
}N\left\langle \Delta x^{i_{1}}\Delta x^{i_{2}}\Delta x^{i_{3}}\Delta
x^{i_{4}}\right\rangle ,
\]
where $\left\langle \Delta x^{i_{1}}\Delta x^{i_{2}}\right\rangle \propto1/N$
and $\left\langle \Delta x^{i_{1}}\Delta x^{i_{2}}\Delta x^{i_{3}}\Delta
x^{i_{4}}\right\rangle \propto1/N^{2}$.

\end{document}